\documentclass[%
 reprint,
superscriptaddress,
 amsmath,amssymb,
 aps,prl
]{revtex4-2}

\usepackage{graphicx}
\usepackage{dcolumn}
\usepackage{bm}

\begin{document}

\preprint{APS/123-QED}

\title{Phase-resolved optical characterization of nanoscale spin waves}

\author{Ond\v{r}ej Wojewoda}
\email{ondrej.wojewoda@vutbr.cz}
\affiliation{%
 CEITEC BUT, Brno University of Technology, Purky\v{n}ova 123, Brno, 612 00, Czech Republic\\
}%
\author{Martin Hrtoň}
\affiliation{%
 CEITEC BUT, Brno University of Technology, Purky\v{n}ova 123, Brno, 612 00, Czech Republic\\
}%
\affiliation{Faculty of Mechanical Engineering, Institute of Physical Engineering, Brno University of Technology, Technick\'{a} 2, Brno, 616 69, Czech Republic}
\author{Meena Dhankhar}
\affiliation{%
 CEITEC BUT, Brno University of Technology, Purky\v{n}ova 123, Brno, 612 00, Czech Republic\\
}%
\author{Jakub Krčma}
\affiliation{Faculty of Mechanical Engineering, Institute of Physical Engineering, Brno University of Technology, Technick\'{a} 2, Brno, 616 69, Czech Republic}
\author{Krist\'{y}na Dav\'{i}dkov\'{a}}
\affiliation{Faculty of Mechanical Engineering, Institute of Physical Engineering, Brno University of Technology, Technick\'{a} 2, Brno, 616 69, Czech Republic}
\author{Jan Kl\'{i}ma}
\affiliation{Faculty of Mechanical Engineering, Institute of Physical Engineering, Brno University of Technology, Technick\'{a} 2, Brno, 616 69, Czech Republic}
\author{Jakub Holobr\'{a}dek}
\affiliation{%
 CEITEC BUT, Brno University of Technology, Purky\v{n}ova 123, Brno, 612 00, Czech Republic\\
}%
\author{Filip Ligmajer}
\affiliation{%
 CEITEC BUT, Brno University of Technology, Purky\v{n}ova 123, Brno, 612 00, Czech Republic\\
}%
\affiliation{Faculty of Mechanical Engineering, Institute of Physical Engineering, Brno University of Technology, Technick\'{a} 2, Brno, 616 69, Czech Republic}
\author{Tom\'{a}\v{s}  \v{S}ikola}
\affiliation{%
 CEITEC BUT, Brno University of Technology, Purky\v{n}ova 123, Brno, 612 00, Czech Republic\\
}%
\affiliation{Faculty of Mechanical Engineering, Institute of Physical Engineering, Brno University of Technology, Technick\'{a} 2, Brno, 616 69, Czech Republic}
\author{Michal Urb\'{a}nek}
\email{michal.urbanek@ceitec.vutbr.cz}
\affiliation{%
 CEITEC BUT, Brno University of Technology, Purky\v{n}ova 123, Brno, 612 00, Czech Republic\\
}%
\affiliation{Faculty of Mechanical Engineering, Institute of Physical Engineering, Brno University of Technology, Technick\'{a} 2, Brno, 616 69, Czech Republic}
\date{\today}

\begin{abstract}
We study theoretically and experimentally the process of Brillouin light scattering on an array of silicon disks on a thin Permalloy layer. We show that phase-resolved Brillouin light scattering microscopy performed on an array of weakly interacting dielectric nanoresonators can detect nanoscale waves and measure their dispersion.  In our experiment, we were able to map the evolution of the phase of the spin wave with a wavelength of 209 nm with a precision of 6 nm. These results demonstrate the feasibility of all-optical phase-resolved characterization of nanoscale spin waves.
\end{abstract}

\maketitle


Spin waves, and their quantum counterpart magnons, are promising candidates for beyond CMOS technology \cite{IRDS:BC2021}. In the framework of spin waves, a lot of ideas for wave-based computing were proposed, such as directional coupler \cite{Wang2020} or inverse-design devices \cite{wang2021inverse, papp2021nanoscale}.  The biggest drawback of these devices is a large group delay caused by an unfavorably low ratio of spin-wave group velocity and wavelength. In recent years, magnonic-research community strives to overcome this drawback by moving towards nanoscale, where the spin waves propagate faster and do not need to travel long distances, thus the group delay is minimized \cite{Chumak2022}. 

When the spin wave wavelengths approach the exchange length of a magnetic material or when dimensions of the magnonic system are reduced, new phenomena such as spin unpinning condition \cite{wang2019spin}, generation of spin waves using parametric pumping \cite{dreyer2022imaging, heinz2022parametric}, or collective dynamics \cite{koerner2022frequency} can arise. Also, the interaction between spin waves and spin textures arises at exchange length scales \cite{che2021confined, Jersch2010, wojewoda2020propagation, yu2013omnidirectional, mayr2021spin}. However, currently the only possibility to image nanoscale spin waves is X-ray microscopy, requiring synchrotron radiation and making investigation of nanoscale-related phenomena very time- and resource-demanding \cite{dieterle2019coherent, albisetti2018nanoscale, sluka2019emission, wintz2016magnetic}. The development of an optical method for spin wave measurement capable to go beyond diffraction limit is one of the major challenges in the field of magneto-optics \cite{kimel20222022}. 


In this letter, we demonstrate that the phase of spin waves can be imaged optically with a subdiffraction resolution and nanometer precision by using Mie resonance-enhanced Brillouin light scattering (BLS) \cite{wojewoda2022observing}. This is achieved by concentrating the light into subdiffractional regions (hotspots) in the magnetic material using high-refractive-index dielectric nanoresonators. The induced spatial restriction allows the light to interact with spin waves that have much shorter wavelengths than the wavelength of free-space light. When we arrange the individual nanoresonators into a weakly-interacting array, it is possible to extract the information about the spin wave at predefined positions. By using standard electron beam lithography (EBL) techniques, the nanoresonators can be placed with nanometer precision.
\begin{figure}
\includegraphics{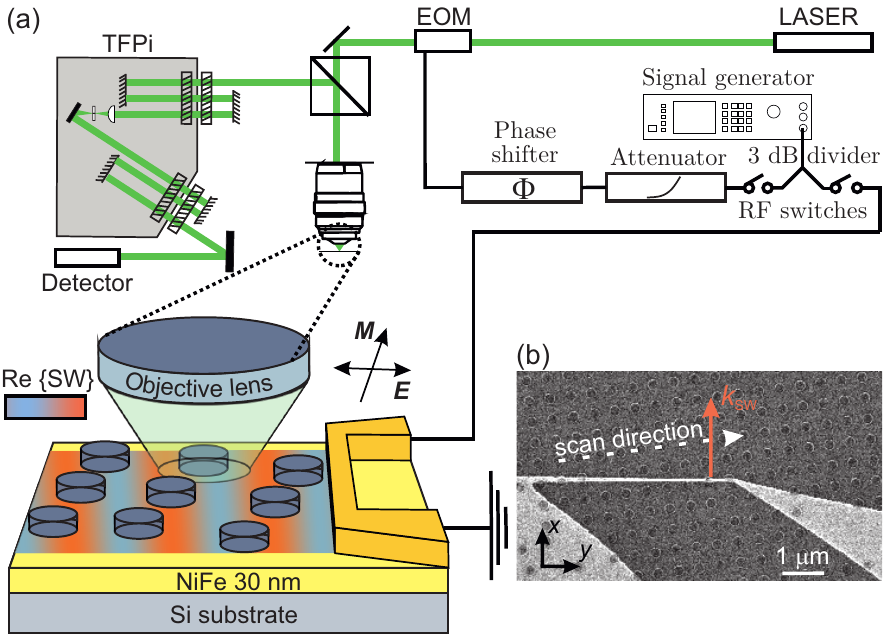}
\caption{\label{fig:Sketch} Geometry of the experiment. (a) Schematics of the used $\mu$-BLS setup. The green light (532\,nm) is generated by a laser and is directed through EOM. Then the light is focused on the sample by an objective lens with NA=0.75. The motorized sample stage allows 2D spatial scanning across individual silicon disks. The backscattered light is analyzed by the Tandem-Fabry-Perot interferometer (TFPi). (b) Scanning electron microscope image of the excitation antenna and silicon disk array. The white dashed line depict scanning line used in Figs. \ref{fig:InterferenceFieldSweep}, \ref{fig:Phase}, and is tilted by 8$^\circ$ with respect to the antenna.}
\end{figure}

To probe the nanoscale spin waves we used a standard phase-resolved micro-BLS setup \cite{flajvsman2022wideband}, see Fig. \ref{fig:Sketch}(a). The light from a single-mode laser (COBOLT Samba) with a wavelength of 532\,nm passes through an electro-optical modulator (EOM, QUBIG TW-15M1-VIS), where the frequency of a tiny fraction of the light is blue- and red-shifted by the pumping frequency. The light is focused on the sample by objective lens with numerical aperture NA=0.75. The backscattered light is analyzed by the Tandem Fabry-Perot interferometer (TFPi, Tablestable, TFP2HC) with a single photon detector \cite{Lindsay1981}.

A single radio-frequency (RF) generator (Rohde \& Schwarz SMB100B) is used to excite the spin waves by an excitation antenna and to drive the EOM at the same time. To ensure a temporal coherence of the EOM and the excited spin waves, the signal is split by a 3\,dB power divider. Two RF switches are used to individually control the sample and EOM excitation. The power of both branches can be independently set by varying the output power of the signal generator and by a variable attenuator which is placed in the EOM branch. A motorized phase shifter, also placed in the EOM branch, allows to precisely set the phase difference between the excited spin waves and the reference signal from the EOM. The excitation antenna is contacted by RF picoprobes (GGB industries, Model 40A). 

The sample consisted of a continuous Permalloy layer (30\,nm-thick) deposited on a silicon substrate. The spin waves were excited by a stripline nanoantenna (180\,nm wide, 105\,nm thick multilayer stack: 10\,nm SiO$_2$/85\,nm Cu/10\,nm Au), fabricated by EBL and lift-off process. The antenna enabled excitation of spin waves with wavelengths down to $\approx 200\,\mathrm{nm}$ $(k \approx 30\,\mathrm{rad}/\mu\mathrm{m})$ \cite{Vavnatka2021, Vlaminck2010}. The square array consisting of 200\,nm-wide and 60\,nm-thick sputter-deposited silicon disks was fabricated by EBL and lift-off process in the vicinity of the antenna. The array had a lattice constant of 500\,nm and a tilt of 8$^\circ$ with respect to the antenna. A scanning electron microscope image of the studied structure is shown in Fig. \ref{fig:Sketch}(b). 

The Brillouin light scattering process is governed by laws of energy (frequency) and momentum (wavenumber) conservation \cite{Kargar2021, Madami2012, Sebastian2015}:
\begin{eqnarray}
    \hbar \omega_\mathrm{i}  =  \hbar ( \omega_\mathrm{s}\ +\omega_\mathrm{m} ) \\
    \hbar \boldsymbol{k}_\mathrm{i}  = \hbar ( \boldsymbol{k}_\mathrm{s}  + \boldsymbol{k}_{\parallel \mathrm{m}} ).
\end{eqnarray}

Here, $\omega$ denotes the frequency, $\boldsymbol{k}$ denotes the wavenumber, $\hbar$ denotes the reduced Planck constant and indices i s, and m stand for incident light, scattered light and magnons, respectively.

Without any complex momentum, in the case of backscattering geometry, these laws of conservation restrict the maximum wavenumber of the spin wave on which the light can be scattered only to the double wavenumber of the probing light \cite{Sebastian2015}. We have recently shown \cite{wojewoda2022observing}, that this limitation can be overcome by concentrating the light into a subdiffraction region, i.e., by introducing the imaginary part of the photon's momentum. This process can be described as an interaction between light and spin wave via magneto-optical coupling mechanism, where the susceptibility ($\chi$) is composed of a static and a dynamic parts \cite{Qiu2000, Cottam1986}. The time- and space-dependent susceptibility can be written as:

\begin{equation}
\hat{\chi}\left(t,\boldsymbol{r}\right)={\hat{\chi}}_{\mathrm{stat}}+Q\left(\begin{matrix}0&iM_z({\boldsymbol{r}},t)&-iM_y({\boldsymbol{r}},t)\\-iM_z({\boldsymbol{r}},t)&0&0\\iM_y({\boldsymbol{r}},t)&0&0\\\end{matrix}\right),
\end{equation}
where ${\hat{\chi}}_\mathrm{stat}$ is the static part,  $M_y$ and $M_z$ are the dynamic components of the magnetization, and $Q$ is the Voigt constant \cite{Qiu2000, Cottam1986}. The inelastic scattering (dynamic part of susceptibility) causes the formation of a polarization current $\boldsymbol{P}$ at a frequency shifted by $\pm\omega_\mathrm{m}$ with respect to the driving field $\boldsymbol{E}$ \cite{Landau1960, Cottam1976, Cottam1986}
\begin{equation}
\boldsymbol{P}\left(\boldsymbol{r},\omega\pm\omega_m\right)=\ \hat{\chi}\left(\boldsymbol{r},\omega_m\right)\ \boldsymbol{E}\left(\boldsymbol{r},\omega\right).
\end{equation}
In our experiment, the momentum ${\boldsymbol{k}}_\mathrm{m}$ and the frequency $\omega_\mathrm{m}$ are defined by the frequency of the RF generator and dispersion relation of spin waves. So in case of monochromatic coherent wave the induced polarization current reads as:
\begin{equation}
\boldsymbol{P}\left(\boldsymbol{r},\omega\pm\omega_m\right)=\ \hat{\chi}\left(\omega_m\right){\ e}^{i{\boldsymbol{k}}_m\cdot\boldsymbol{r}}\ \boldsymbol{E}\left(\boldsymbol{r},\omega\right).    
\label{eq:InducedPolarization}
\end{equation}
The above equation highlights the importance of field localization provided by the silicon disks: the spatial profile of $\boldsymbol{E}$ determines the area that contributes to the collected BLS signal. If the modulation of the spatial profile of $\boldsymbol{E}$ is larger than the spin-wave wavelength, the exponential factor ${\ e}^{i{\boldsymbol{k}}_m\cdot\boldsymbol{r}}$ is averaged out and the information about the spin wave is lost \cite{wojewoda2022observing, Freeman2020}. On the other hand, intense hotspots generated by the silicon disk effectively limit the spatial extent of the induced polarization to a very small volume and thus facilitate the extraction of the spin-wave phase from BLS measurements. 

To confirm these findings, we performed a 2D scan of the BLS intensity in the vicinity of the excitation antenna which was connected to the RF generator with frequency set to 14.5\,GHz, in an external magnetic field of 50\,mT. The antenna thus excited coherent spin waves with wavevector $k=27\,\mathrm{rad}/\mu\mathrm{m}$ ($\lambda=232$\,nm) \cite{Kalinikos1986, githubSWT}. Such short-wavelength spin waves are beyond the detection limit of conventional $\mu$-BLS (the detection limit of our $\mu$-BLS is $k \approx 11\,\mathrm{rad}/\mu\mathrm{m})$ \cite{wojewoda2022observing}, thus without any further enhancement we would see no signal. However, due to the subdiffraction localization of the $E$-field by the silicon disks, we can observe a propagating spin wave with even such a short wavelength (see Fig. 2). The BLS signal naturally decays as the wave propagates further from the excitation antenna and it is strongly enhanced at the positions of the silicon disks.
This agrees well with expectations based on Eq. (\ref{eq:InducedPolarization}).

\begin{figure}
\includegraphics{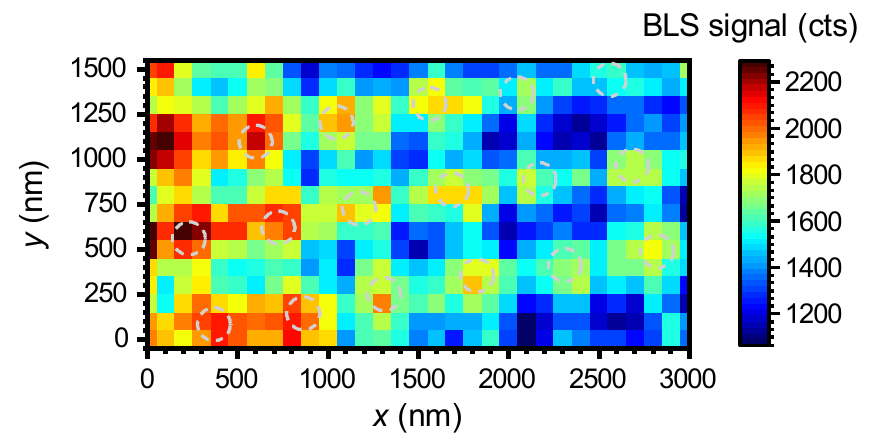}
\caption{\label{fig:BLS2Dmap} Spatial map of BLS intensity at the excitation frequency of 14.5\,GHz. At this frequency the spin-wave wavelength is 232\,nm; such a short wavelength is not possible to detect with conventional $\mu$-BLS (without the presence of the silicon disk array). The positions of the individual silicon disks are marked by dashed line.}
\end{figure}

In the second experiment, we focused on the measurement of the spatial evolution of the spin wave phase. During the BLS process the scattered photon acquires the phase of the spin wave and this phase can be reconstructed with the help of reference phase signal produced by the EOM. In this measurement both RF switches were switched on [see Fig. \ref{fig:Sketch}(a)], and we observed the interference of the light inelastically scattered on spin-waves with the light modulated by EOM \cite{vogt2009all}. In order to acquire the phase with spatial resolution sufficient to measure spin waves with very short wavelengths, we exploited the symmetry along $y$-axis of the spin wave propagating perpendicularly from the excitation antenna and performed a linescan over one row of the silicon disk array [see Fig. \ref{fig:Sketch}(b)]. As the array is tilted by 8°, the distance of the disks from the antenna increases by 70\,nm for each disk in the row.

\begin{figure}
\includegraphics{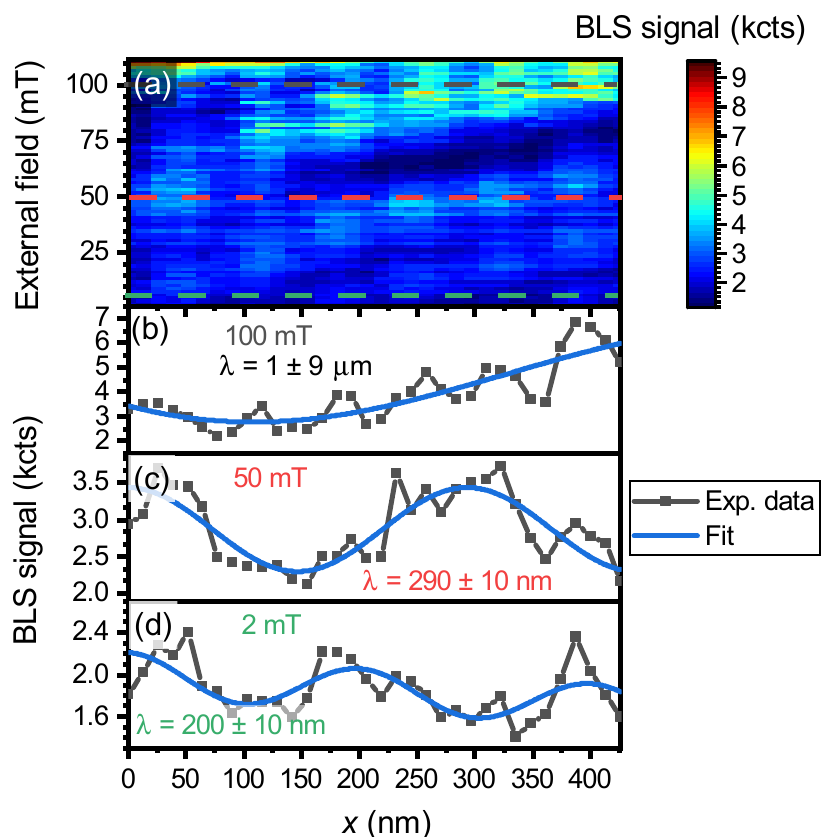}
\caption{\label{fig:InterferenceFieldSweep} Interference BLS measurement at 14\,GHz.  (a) The BLS intensity along the line tilted by 8$^\circ$ with respect to the excitation antenna in different external fields. (b), (c), (d) Slices of the data presented in (a) for 100\,mT (b), 50\,mT (c), and 2\,mT (d) are shown together with the corresponding fits by the two-wave interference model.}
\end{figure}

The measured dependence of the interference signal ($I$) on the distance from the excitation antenna and on the external magnetic field is plotted in Fig. \ref{fig:InterferenceFieldSweep}(a). The cross sections of this plot at magnetic field values of 100\,mT, 50\,mT, and 2\,mT are shown in Fig. \ref{fig:InterferenceFieldSweep} (b), (c) and (d), respectively. We can observe how the spin wave wavelengths shorten for the lower values of the external magnetic field. The shortening is apparent from the gradual change of the distance between the interference minima and maxima. To obtain the exact value of the spin wave wavelength, we fit the data with a two-wave interference model \cite{flajvsman2020zero, vogt2009all}:
\begin{equation}
I = R_0\exp\left(\frac{x}{\delta}\right)+E+\sqrt{E{\ R}_0\exp\left(\frac{x}{\delta}\right)}\cos\left(\frac{2\pi x}{\lambda}\ +\ \phi_0\right),
\end{equation}
where $R_0$ and $E$ are parameters describing the signal intensity from the spin wave and the EOM, respectively, $\lambda$ is the spin-wave wavelength, $\delta$ is the spin-wave decay length, and $\phi_0$ is the initial phase offset. The fitted wavelengths are in corresponding panels of Fig. \ref{fig:InterferenceFieldSweep}.
\begin{figure*}
\includegraphics{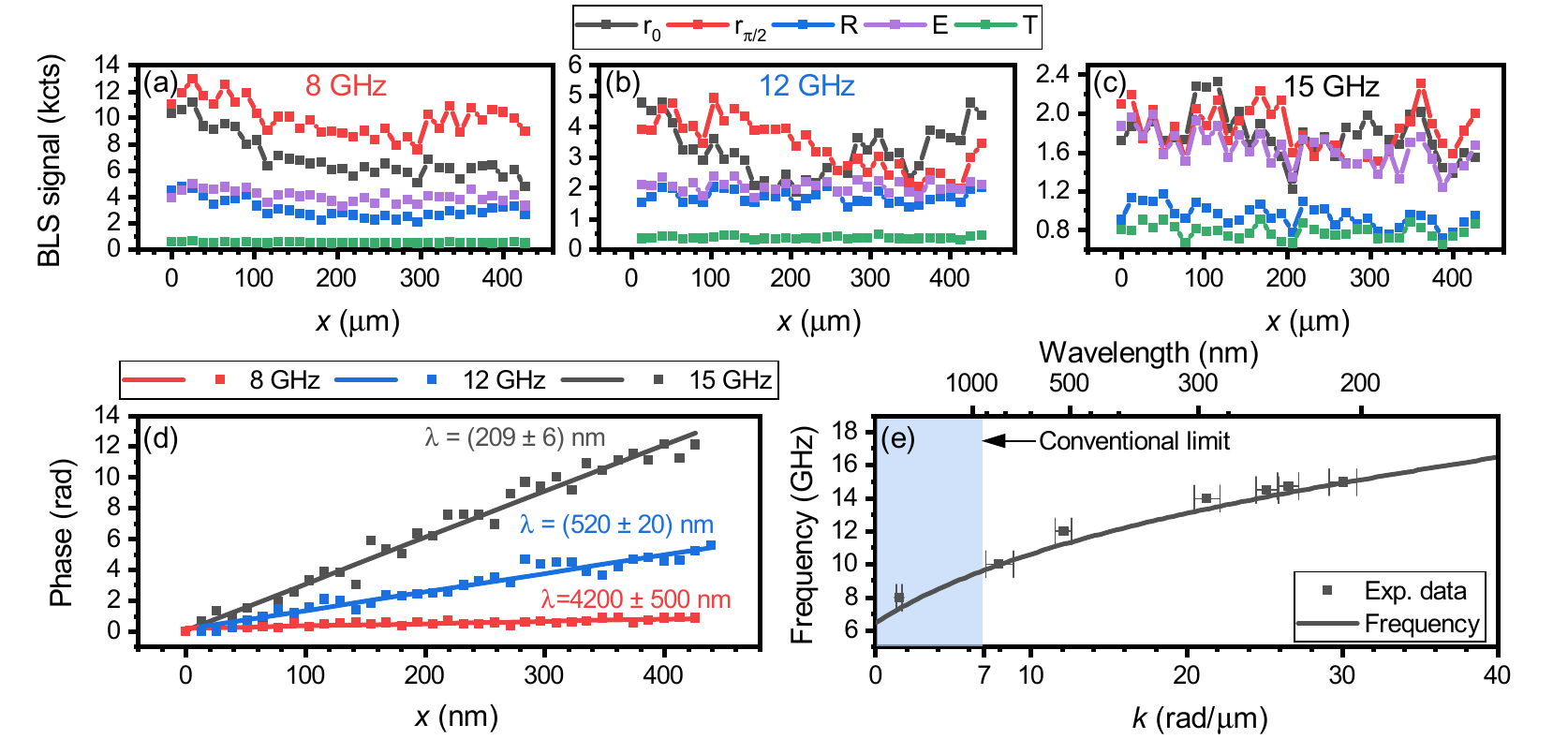}
\caption{\label{fig:Phase} Extraction of the spin-wave dispersion relation from phase-resolved BLS measurement. (a), (b), (c), Five signals needed for the full phase extraction  at 8, 12, 15\,GHz, respectively. (d) Evolution of the phase versus the coordinate in the direction of propagation of the spin waves at 8, 12, 15\,GHz. (e) Extracted dispersion (squares) and analytical calculation (solid line) \cite{githubSWT, Kalinikos1986}. The blue rectangle depict the limit of conventional $\mu$-BLS.}
\end{figure*}

Although it was possible to obtain spin-wave wavelengths from the BLS interference signal and nonlinear fitting with the two-wave interference model, the technique of the full-phase reconstruction \cite{serga2006phase, bouchal2019high} proved to be much more robust and we used this technique to obtain the full spin-wave dispersion over a broad range of frequencies and $k$-vectors. This technique requires four measurements to fully reconstruct the spin-wave phase. In order to account for thermally excited spin waves, we added fifth measurement, which must be taken into account if the coherent signal is comparable to the incoherent thermal background. To calculate the evolution of the spin-wave phase along the distance $x$, real $\mathrm{Re}[\Psi\left(x\right)]$ and imaginary $\mathrm{Im}[\Psi\left(x\right)]$ parts of the wave amplitude have to be calculated:
\begin{eqnarray}
    \mathrm{Re}[\Psi\left(x\right)]=\frac{r_0-R-E+T}{2\sqrt{E-T}},\\
    \mathrm{Im}[\Psi\left(x\right)]=\frac{r_{\pi/2}-R-E+T}{2\sqrt{E-T}}.
\end{eqnarray}
Here, $r_0\ \left(r_{\pi/2}\right)$ is the interference signal with EOM phase shifted by 0 ($\pi/2$), $R$ is the signal from coherent spin waves (which includes also a contribution from thermal spin waves), $E$ is the EOM signal, and $T$ is a signal from thermal spin waves. These signals are shown in Figs. \ref{fig:Phase}(a), (b), and (c) for 8, 12, and 15\,GHz, respectively. The spin-wave phase can be calculated as $\Theta \left(x\right)=\mathrm{atan} \left(\frac{\mathrm{Re} \left(\Psi\right)}{\mathrm{Im} \left(\Psi\right) } \right)$. In the case of a monochromatic spin wave, the dependence $\Theta\left(x\right)$ is linear and its slope directly represents the spin-wave wavenumber. The phases reconstructed from the experimental data in Fig. \ref{fig:Phase}(a,b,c) are shown in Fig \ref{fig:Phase}d, together with linear fits of their slopes. 

The wavevectors extracted from the phase profiles acquired in the field of 50\,mT at various frequencies are compared to the theoretical dispersion relation [see Fig. \ref{fig:Phase}(e)] calculated according to the Kalinikos-Slavin model \cite{Kalinikos1986, githubSWT}:
\begin{equation}
\omega^2=\left(\omega_\mathrm{H}+A^2\omega_\mathrm{M} k^2\right)\left(\omega_\mathrm{H}+A^2\omega_\mathrm{M} k^2+\omega_\mathrm{M} F_n\right),
\label{eq:KalSlav}
\end{equation}
where $\omega=2 \pi f$ is the spin-wave frequency, $k$ is its wavevector, $\omega_\mathrm{H}=\mu_0\gamma H_\mathrm{ext}$, $\omega_\mathrm{M}=\mu_0\gamma M_\mathrm{s}$, $M_\mathrm{s} = 741\,\mathrm{kA/m}$ is the saturation magnetization,  $\gamma  = 29.5\,$GHz/T is the gyromagnetic ratio, $\mu_0$ is the permeability of a vacuum, $A = 6.81$\,nm is the exchange length, $F_n=1+\frac{\left(1-e^{-kt}\right)\left(1-\frac{1-e^{-kt}}{kt}\right)\omega_M}{kt(\omega_H+Ak^2\omega_M)}$, and $t = 34.8$\,nm, is the thickness of the sample \cite{wojewoda2022observing}. 
The excellent agreement between the theory and the experimental data [see Fig. \ref{fig:Phase}(e)] for wavevectors ranging from 4\,rad/$\mu$m up to 30\,rad/$\mu$m means that we were able to reliably measure and reconstruct the spin wave phase even for spin-wave wavelengths where the BLS process had to be mediated by the silicon disk. The maximum measured wavevector corresponds to the excitation limit of the 180\,nm-wide antenna \cite{Vavnatka2021, Vlaminck2010}. We anticipate that with another type of a spin wave source \cite{baumgaertl2020nanoimaging, mayr2021spin, urazhdin2014nanomagnonic}, spin waves with even shorter wavelengths could be measured. 

When we look more carefully at the linescan data we can observe a periodic pattern [either areas with an increased signal nicely visible in Fig \ref{fig:InterferenceFieldSweep}(b) or stair-like pattern in Fig \ref{fig:Phase}(d)]. This pattern has a periodicity of 70\,nm, which is the same as periodicity of the positions of the individual silicon disks under the scanning tilt. The linescans were acquired with oversampling, where the spatial step in the spin-waves propagation direction ($x$ axis) was approx. 13\,nm, while the distance between the silicon disks was approx. 70\,nm. This periodic pattern in extracted phase [Fig. \ref{fig:Phase}(e)] suggests that the detected phase is defined by the positions of the disks on the thin film and is not overly sensitive to the exact positioning of the laser spot relative to the position of the nanoresonator. Also note, that we were able to reliably measure the spin-wave wavelength  $\lambda$ = 4.2\,$\mu$m by scanning over a distance of 450\,nm only. This would not be possible without precisely positioned detection points and without the full-phase reconstruction technique.

\begin{figure}
\includegraphics{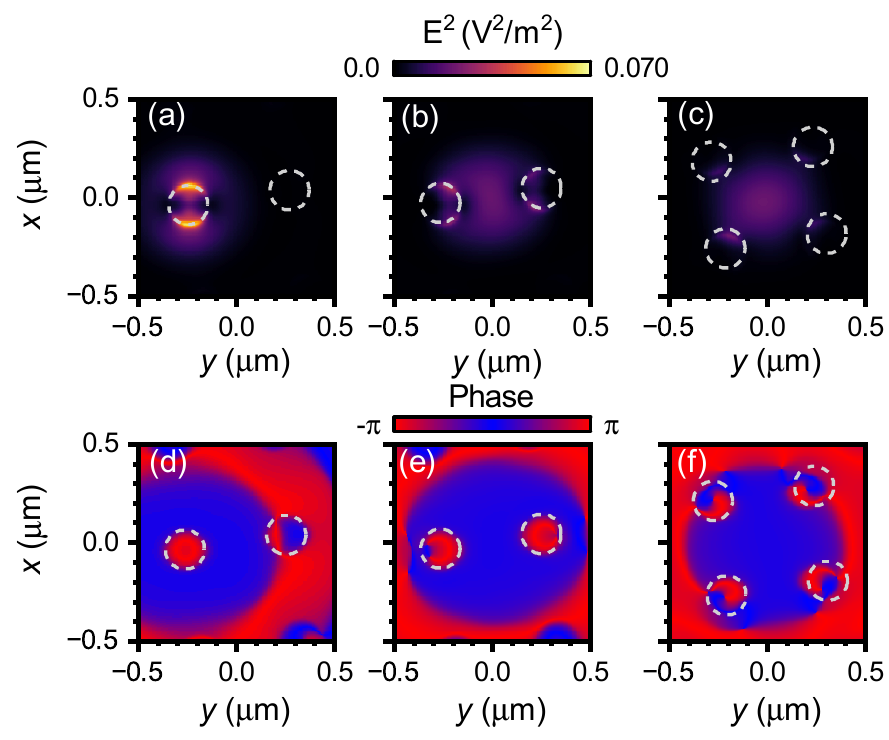}
\caption{\label{fig:Simul} Scanning with a weakly-interacting array of silicon disks. Finite-differences-time-domain simulation of the distribution of the squared electric intensity [(a), (b), (c)] and phase [(d), (e), (f)]  in the plane of the Permalloy layer. For the precisely positioned beam on the silicon disk in the array (a), (d), case in between two disks (b), (e) and the case when the beam hits the exact middle of the array (c), (f). The gray dashed line depicts the positions of silicon disks.}
\end{figure}

To clarify the phase stability when scanning over the array of silicon disks, we performed finite-differences-time-domain (FDTD) simulations. of the sample with the same parameters as in the experiments. 
The simulations were carried out in Lumerical software with the mesh cell set to 5\,nm in all directions. More details can be found in \cite{wojewoda2022observing}. Figure \ref{fig:Simul}(a)  shows the localization of electromagnetic field into hotspots, that allow the detection of high-$k$ spin waves. In this case the Gaussian beam is precisely positioned on the silicon disk. On the other hand, if we position the beam spot directly between the silicon disks [Fig. \ref{fig:Simul} (b-c)], this localization is absent, and the overall light intensity is lower. This results in a decrease of the detection sensitivity to high-$k$ spin waves and overall decrease of the BLS signal. We also extracted the  phase of the electric field from the simulated data [Fig. \ref{fig:Simul}(d-f)]. The phase in the axis of the linear polarization of the incident light ($x$) is homogeneous across the high intensity regions (hotspots), while the inner area of the disk has the opposite phase and this is the same for all simulated positions of the beam. This independence of the electric-field phase on exact position of the beam suggests the robustness of the phase reconstruction to, e.g., mechanical vibrations.
Note that even though the two hotspots are localized on the disk edges and separated by 200\,nm, this does not influence the capability of the system to detect even nanometer changes in the spin-wave wavelength. This is analogous to the case of inductive detection of spin waves by coplanar waveguides or meander antennas in propagating spin-wave spectroscopy experiments. There, the electromagnetic field distribution under relatively large detection antennas with complex geometries is also non-trivial, and still the spin wave phase can be measured \cite{Vavnatka2021, lucassen2019optimizing}.   


In conclusion, we demonstrated an efficient and simple technique for phase-resolved characterization of nanoscale spin waves based on conventional phase-resolved BLS microscopy. Presented technique uses weakly interacting arrays of Mie resonators which localize and enhance the electric field involved in the magneto-optical coupling and the connected BLS process. Short-wavelength spin waves can be detected, and their phase can be characterized with nanometer precision. This precision depends on the exact positioning of the nanoresonator in the array. With modern electron beam lithography techniques, sub-10 nm precision of placement of individual array elements is easily achievable. The presented experiments were limited only by the excitation efficiency of the used stripline nanoantenna. With the use of a different type of the spin-wave source it should be possible to detect spin waves with sub-100\,nm wavelengths. Also, with the presented technique other types of BLS experiments, such as time-resolved measurements \cite{merbouche2022giant}  can be easily performed on nanoscale spin waves. These results show that Mie-enhanced BLS microscopy is relevant and highly versatile tool for nanoscale magnonics research.

\begin{acknowledgements}
We gratefully acknowledge CzechNanoLab project LM2018110 for the financial support of the measurements and sample fabrication at CEITEC Nano Research Infrastructure. O.W. was supported by Brno PhD talent scholarship and acknowledges BUT specific research project. F.L. acknowledges the support by the Grant Agency of the Czech Republic (21-29468S). J.H. acknowledges a support from the project Quality Internal Grants of BUT (KInG BUT), Reg. No. CZ.02.2.69 / 0.0 / 0.0 / 19\_073 / 0016948, which is financed from the OP RDE.
\end{acknowledgements}
\bibliography{apssamp}

\end{document}